\documentclass{iopart}
\usepackage{graphicx}    
 \usepackage{amssymb}
 \usepackage[square,sort&compress]{natbib}

\begin{document}
\title[IR luminescence of Xe$_{2}$ excimers]{Low- and high-density features of IR luminescence of Xe$_{2}$ excimers produced by electron impact}

\author{A. F. Borghesani$^{1}$, G. Carugno$^{2}$, and I. Mogentale$^{1}$}
\address{$^{1}$ Department of Physics, University of Padua, Italy}
\address{$^{2}$ I.N.F.N., Sezione di Padova, via F. Marzolo 8, I--35131 Padua, Italy}
\ead{borghesani@padova.infm.it}
\begin{abstract}
Electron--impact excitation of Xe atoms in pure Xe gas and in a Xe(10 \%)--Ar(90 \%) mixture has led to the discovery of infrared (IR) luminescence of Xe$_{2}$ excimers. The investigation of the emission spectrum at low gas density has allowed the identification of the molecular states involved in the transition. When the gas density is increased to values up to 40 times larger than the density of the ideal gas at standard temperature and pressure, the interaction of the excimer with the dense environment produces a strong red--shift of the spectrum that is interpreted in terms of many--body effects.\end{abstract}
\submitto{Physica Scripta}
\maketitle
\section{Introduction}

The luminescence of rare gas excimers is studied extensively for applications in the vacuum ultraviolet (VUV) including production of coherent and incoherent VUV light sources \cite{ledru2006} and detection of ionizing radiation \cite{knoll}.
Xe is of particular interest because of its efficiency to convert the excitation energy in intense VUV radiation. When the gas is excited by means of several techniques including electrical discharges \cite{colli1954}, irradiation with ionizing particles \cite{arai1969,koe1974,davide1}, or resonance lines \cite{broadmann1976,bonifield1980,dehmer1986,dutuit1978,gornik1982}, three--body collisions of excited and ground state atoms lead to the formation of the excited dimer, i. e., excimer, Xe$_{2}^{\star}.$ 
The decay of the lowest lying excited levels towards the dissociative ground state of the molecule gives origin to the emission of the 1st and 2nd VUV continua at about 152 and 170 nm, respectively. 

Owing to their relevance for so many applications, these VUV excimer bands have been thouroughly studied theoretically  \cite{mulliken1970,ermler1978,jonin2002_I,jonin2002_II} and experimentally \cite{gornik1982,keto1974,castex1981,raymond1984,koe1995} in order to ascertain the molecular structure and potential energy curves and the kinetics of the deexcitation processes leading to the population of the $6s$ excited atomic states that are eventually involved in the production of these molecular states \cite{ledru2006,moutard1988,leichner1976,wenck1979,museur1994,millet1978,broadmann1977,bonifield1980,salamero1984,moutard1986,alekseev1999}.

At quite low gas pressure ($P< 20$ kPa), the VUV luminesce signal consists of the 1st continuum. On the contrary, at higher pressure ($P>50 $ kPa), only the 2nd continuum is present in the spectrum. This different behavior of the two continua as a function of pressure has been explained by time--resolved spectroscopic studies showing that the 1st continuum is due to radiative transitions from the vibrationally excited $(0_{u}^{+})_{v^{\prime} \gg 0}$ state correlated with the resonant $5p^{5}6s\,(^{3}\!P_{1})$ atomic state towards the dissociative $0_{g}^{+}$ ground state, whereas the 2nd continuum consists of the overlapping bound--free emission from the lowest vibrationally relaxed $(0_{u}^{-},1_{u})$ states correlated with the metastable $5p^{5}6s \,(^{3}\!P_{2})$ atomic state.                                

Nearly no attention at all has been paid to the possibility that molecular infrared (IR) transitions may occur
in the cascade of processes leading to the formation of the excimer states radiating in the VUV. Only an IR spectrum centered about 800 nm has been observed and its appearance has been attributed to a $0_{g}^{+}\,(6p[1/2]_{0})\rightarrow 0_{u}^{+}\,(6s [3/2]_{1})_{v^{\prime}\gg 0}$ transition \cite{dehmer1986}. One possible reason for this lack of interest on IR processes may be that the minimum of higher--lying excimer states occurs at an internuclear distance at which the weakly bound molecular ground state is strongly repulsive and is therefore not so easily accessible by multiphoton selective excitation.  On the contrary, broad--band excitation techniques using high--energy charged particles \cite{leichner1976,davide1} may produce excited atoms of such high kinetic energy that may collide with ground state atoms at short distance yielding molecular states in higher energy levels although their parity cannot be controlled. 

We have recently discovered a broad band in the near IR region centered at a wave number $\tilde \nu_{m} \approx 7800$ cm$^{-1},$ i. e., $\lambda \approx 1.3 \,\mu$m. The emission is produced by exciting the gas at room temperature by means of short pulses of electrons of 70 keV energy \cite{borghesani2001,borghesani2005}. Technical details of the experiment can be found in literature. 

The IR band is quite broad. At $P\approx 20$ kPa, its FWHM is $\Gamma\approx 900$ cm$^{-1}.$ The relative width $\Gamma/\tilde\nu_{m}\approx 0.115$ is comparable with that of the 2nd continuum \cite{koe1974,jonin1998}. This IR band occurs both in the pure gas and in an Ar (90 \%)--Xe (10 \%) mixture. As the features of the spectrum in both gases in the limit of low $P$ are identical, we attribute the emission to a Xe$_{2}$ molecular bound--free transition. 

Improvements in our experimental apparatus have allowed us to carry out more accurate time--integrated measurements of the excimer IR emission at low pressure
and also to extend the investigated density range up to density nearly 40 times larger than the density of the ideal gas at standard temperature and pressure (273 K and 0.1 MPa). By so doing, we have now the opportunity to better assign the molecular states involved in the transition and to elucidate the interaction of the excited molecules with a dense gas environment. 

The electronic structure of homonuclear rare gas excimers can be described by an ionic molecular core surrounded by an electron in a diffuse Rydberg orbital much larger in diameter than the internuclear distance \cite{mulliken1970,arai1978,audouard1991}. Such a state can still exist in a high--pressure environment provided that the optically active electron is only weakly scattered off the atoms of the host gas. This occurs  if the electron mean free path is larger than the orbit of the Rydberg state as it happens to be at low density.  It is therefore interesting to look at the evolution of a molecular Rydberg state when the environment becomes increasingly denser. In our previous experiments \cite{borghesani2001,borghesani2005} we have mainly detected a red--shift of the IR emission spectrum of the excimer. In this experiment we confirm the previous results in an extended density range.

\section{Low--density results}

\begin{figure}[htbp]
 \begin{center}
 \includegraphics[scale=0.5]{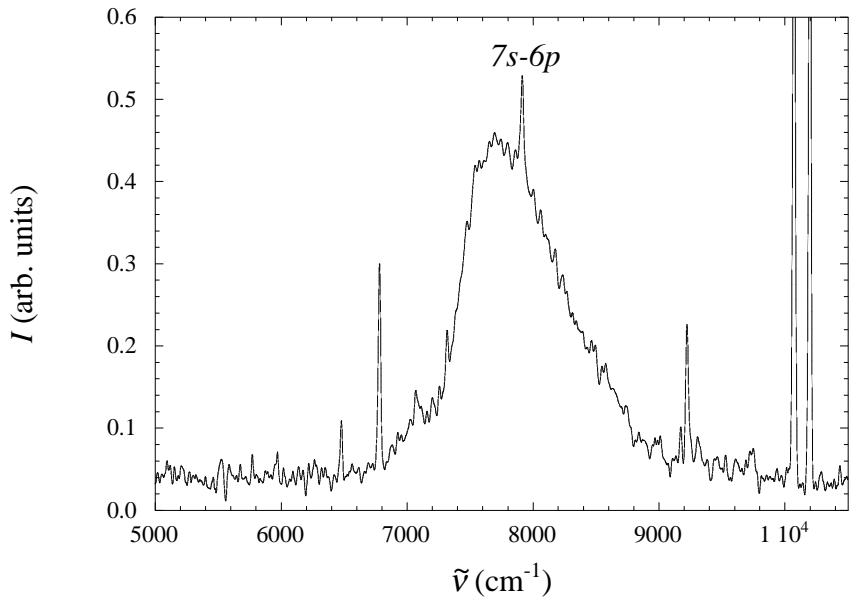}
\caption{High--resolution excimer IR emission spectrum. The atomic line related to a $7s-6p$ transition is shown. \label{fig:eccimeropuro}}\end{center} 
 \end{figure}

In Fig. \ref{fig:eccimeropuro} we show an experimental IR emission spectrum recorded in pure Xe gas at $P\approx 0.1 $ MPa at room temperature in an extended wave number range with a resolution of 16 cm$^{-1}.$
The broad line shape is the molecular excimer band centered at $\tilde \nu_{m}\approx 7800$ cm$^{-1}.$ 

Several atomic lines appear in the emission spectrum as to be expected because of the broad band excitation induced by electron impact. 
It is particularly interesting to note that an atomic line produced in a $7s\,[1/2]^{o}-6p\,[1/2]_{0}$ atomic transition \cite{NIST} appears to be embedded in the excimer band. The closeness of such an atomic transition to the excimer spectrum suggests that the upper molecular state may be correlated with a $5p^{5} 6p$ atomic limit.
 
Only recently theoretical potential energy curves for the higher lying molecular states of Xe$_{2}$ have appeared in literature \cite{jonin2002_I,jonin2002_II}. It is therefore now possible to test the accuracy of some of the theoretical potentials by comparing this 
first experimental excimer IR emission spectrum with that simulated by using standard Franck--Condon calculations \cite{herzberg,telli}.

We thus assume that a bound--free transition is responsible of the observed emission and that the upper bound state is correlated with the $6p\,[1/2]_{0}$ atomic state. The molecular potential correlated with this state shows an avoided crossing with the $g$ molecular state related to the $5d\,[1/2]_{1}$ state leading to predissociation \cite{museur1994}. The predissociation lifetime is estimated to be $\tau_{p}\approx 10^{-10}$ s, whereas the mean time between collisions at the density of the present experiment, $N_{e}\approx 2.4\cdot 10^{25}$ m$^{-3},$ is $\tau_{c}\leq10^{-11}$ s. Thus, the large collision rate electronically stabilizes excimers before they predissociate. 

At present, there are no estimates of the radiative lifetime and of the decay rate for vibrational relaxation for this bound state. We thus assume that they are not too different from those of the states responsible for the emission of the VUV continua. The radiative lifetimes of the highly excited vibrational states of the $0_{u}^{+}$ and $(1_{u}, 0_{u}^{-})$  molecular states correlated with the $6s$ atomic limit are estimated to be $\tau_{1}\approx 5$ ns \cite{moutard1988} and $\tau_{2}\approx 40$ ns \cite{keto1974}, respectively. Similarly, the decay rate for vibrational relaxation $k_{3} $ is the same for all those state \cite{bonifield1980}, yielding a decay time $\tau_{3} =(k_{3}N_{e})^{-1} \approx 0.65 $ ns. 

This hierarchy of characteristic times $\tau_{c}<\tau_{p}<\tau_{3}<\tau_{1}<\tau_{2}$ supports the assumption that collisions stabilize excimers electronically and quickly establish thermal equilibrium among their internal rovibrational degrees of freedom.

Franck--Condon calculations require the knowledge of the potential energy curves whose choice among those reported in literature \cite{jonin2002_II} has to comply with several requirements: 1) the upper bound state must belong to the $6p$ atomic manifold; 2) the parity of the initial and final state must obey the selection rules $u\leftrightarrow g$ and $^{+}\!\leftrightarrow^{+};$ 3) the energy difference between the two potential energy curves at the coordinate of the minimum of the bound state potential must approximately be equal to the central wave number $\tilde \nu_{m}$ of the observed spectrum.

According to these criteria, the upper bound state has been assigned as the $(3)0_{u}^{+} $ ungerade state correlated with the $6p\,(^{1}\!D_{2})$ limit (with some contributions due to the mixing with a $5d$ atomic state) \cite{jonin2002_II}. The lower dissociative state has been assigned as the $(1)0_{g}^{+}$ gerade state correlated with the $6s\,(^{3}\!P_{1})$ resonant atomic state \cite{jonin2002_II}. 
For both of these molecular states the projection $\Omega$ of the electronic angular momentum along the molecular axis is zero. Thus, as
in this transition $\Omega = 0\rightarrow \Omega =0,$ the additional selection rule for the total molecular angular momentum $\Delta J =\pm 1 $ must be also taken into account.

The intensity of the line shape, at infinite resolution, is computed within the centroid approximation because the electronic transition moments are not known as a function of the internuclear coordinates for the states at hand \cite{herzberg,telli}, thus yielding
\begin{eqnarray}
 I\left( \tilde \nu\right)&\propto &\sum\limits_{v^{\prime}J^{\prime}} e^{-\beta E_{v^{\prime}J^{\prime}}}
\tilde\nu^{4} \left\{ \left(J^{\prime} +1\right) \Big\vert \langle \epsilon^{\prime\prime}, J^{\prime}+1\vert v^{\prime},J^{\prime}\rangle \Big\vert^{2} + \right.
 \nonumber\\ 
 &+ & \left. J^{\prime}  \Big\vert \langle \epsilon^{\prime\prime}, J^{\prime}-1\vert v^{\prime},J^{\prime}\rangle \Big\vert^{2} 
 \right\}\label{eq:FCspc}\end{eqnarray}
in which the selection rule $J^{\prime}-J^{\prime\prime} =\pm 1$ is enforced. As usual, a single prime refers to the initial state (the bound one, in this case) and a double prime refers to the final state \cite{herzberg}. 
$E_{v^{\prime}J^{\prime}} $ is the energy of a rovibrational level of the bound state and is given by
\begin{equation}
E_{v^{\prime}J^{\prime}}  = D_{e}^{\prime}\left[1 +\epsilon_{v^{\prime}}\right] + B_{e}^{\prime} J^{\prime}\left(J^{\prime}+1\right)
\label{eq:roviben}\end{equation}
where $D_{e}^{\prime} $ is the dissociation energy of the upper state and $\epsilon_{v^{\prime}}$ are the scaled vibrational eigenvalues measured from the dissociation limit.
$B_{e}^{\prime} = \hbar^{2}/2m_{r}R_{e^{\prime}}$ is the rotational constant, in which $m_{r}$ is the Xe$_{2}$ reduced mass and $ R_{e^{\prime}}$ is the equilibrium internuclear distance of the bound state.
The energy of the emitted photon is
\begin{equation}
\tilde \nu = \left[T_{e}^{\prime}-T_{e}^{\prime\prime} - D_{e}^{\prime\prime}
+  E_{v^{\prime}J^{\prime}} -\epsilon^{\prime\prime} \right]
\label{eq:wn}\end{equation}
in which $T_{e}^{\prime}$ and $T_{e}^{\prime\prime}$
 are the values of the minimum of the potential energy curves for the upper and lower states and $D_{e}^{\prime}$ and $D_{e}^{\prime\prime}$ are their well depth. It has to be recalled that the $0_{g}^{+}$ state, though mainly dissociative, has a small attractive well at quite large internuclear distance, with the minimum located in $R_{e^{\prime\prime}}.$
 
  $\vert v^{\prime},J^{\prime}\rangle $ is the eigenfunction of a rovibrational state of the bound potential. $\vert\epsilon^{\prime\prime},J^{\prime\prime}\rangle$ is a scattering state of kinetic energy $\epsilon^{\prime\prime}$ and total angular momentum $J^{\prime\prime}$ in the vibrational continuum of the dissociative potential.
 
 The Boltzmann factor in Eq. \ref{eq:FCspc} accounts for the equilibrium thermal distribution of the rovibrational degrees of freedom with $\beta^{-1} \approx 208.5$ cm$^{-1}$ at room temperature. 
 
The vibrational eigenvalues $\epsilon_{v^{\prime}} $ and eigenfunctions $\vert v^{\prime},J^{\prime}\rangle $ are found by integrating numerically the Schr\H odinger equation for the rotationless potential of the bound state by adopting the Numerov--Cooley finite difference scheme \cite{Koonin} and replacing the centrifugal potential with the constant value $B_{e}^{\prime}J^{\prime}(J^{\prime}+1)$ \cite{hw}. For numerical purposes, the literature $0_{u}^{+}$ potential has been accurately fitted to a Morse potential
\begin{equation}
V_{0_{u}^{+}} (R) = T_{e}^{\prime} +D_{e}^{\prime} \left\{
1 -e^{\left[ - \beta_{e^{\prime}} \left(R -R_{e^{\prime}}\right) \right]
}\right\}^{2}
\label{eq:morse}\end{equation}
with $T_{e}^{\prime}= 13860 $ cm$^{-1},$ $D_{e}^{\prime} = 1717$ cm $^{-1},$ $R_{e^{\prime}}= 3.23$ \AA, and $\beta_{e^{\prime}}R_{e^{\prime}}= 6.734.$ The bound state accomodates approximately 34 vibrational states though only the first few contribute significantly to the spectrum because of the Boltzmann factor.

The numerical value of the rotational constant is quite small $B_{e}^{\prime}\approx 2.47 \cdot 10^{-2}$~cm$^{-1},$ corresponding to a rotational temperature $\Theta_{r}\approx 3.5\cdot 10^{-2}$ K. Thus, the thermal population of rotational states is non negligible  up to $J^{\prime}\approx 250.$

The scattering eigenfunctions of the dissociative potential $\vert \epsilon^{\prime\prime},J^{\prime\prime}\rangle$ are found by numerically integrating the Schr\H odinger equation with the effective potential
\begin{equation}
V_{f}= V_{0_{g}^{+}}(R) +\frac{\hbar^{2}}{2m_{r}R^{2}}J^{\prime\prime}\left(J^{\prime\prime}+1\right)
\label{eq:veff}\end{equation}
in which $ V_{0_{g}^{+}}(R) $ is the rotationless potential of the dissociative state.
$ V_{0_{g}^{+}}(R) $ is characterized by a very shallow well of depth $D_{e}^{\prime\prime}\approx 218$ cm $^{-1}$ occurring at $R_{e^{\prime\prime}}\approx 4.92$~\AA\ and by a value $T_{e}^{\prime\prime}\approx 4779$ cm$^{-1}$ \cite{jonin2002_II}. For numerical purposes, the potential is cast in the form $ V_{0_{g}^{+}}(R) = T_{e}^{\prime\prime} +D_{e}^{\prime\prime}f(x),$ in which $f(x)$ is an analytical HFD--B potential \cite{aziz1986} and $x=R/R_{e^{\prime\prime}}$. 
A Runge--Kutta 4-th order scheme with adaptive stepsize control has been adopted to integrate the Schr\H odinger equation \cite{NR}. 
The scattering wave functions are normalized to unitary incoming flux \cite{gasio}
\begin{equation}
\psi_{\epsilon^{\prime\prime}} \equiv \langle R\vert\epsilon^{\prime\prime}, J^{\prime\prime}\rangle 
\stackrel{R\rightarrow\infty}{\longrightarrow} 
\left(
\frac{2m_{r}}{\pi\hbar^{2}k}\right)^{1/2}\sin{\left(kR+\eta\right)}
\
\label{eq:}\end{equation}
in which $k= [(2m_{r}/\hbar^{2})\epsilon^{\prime\prime}]^{1/2}$ is the wave vector and $\eta$ is the appropriate phaseshift.

\begin{figure}[htbp]
 \begin{center}
 \includegraphics[scale=0.5]{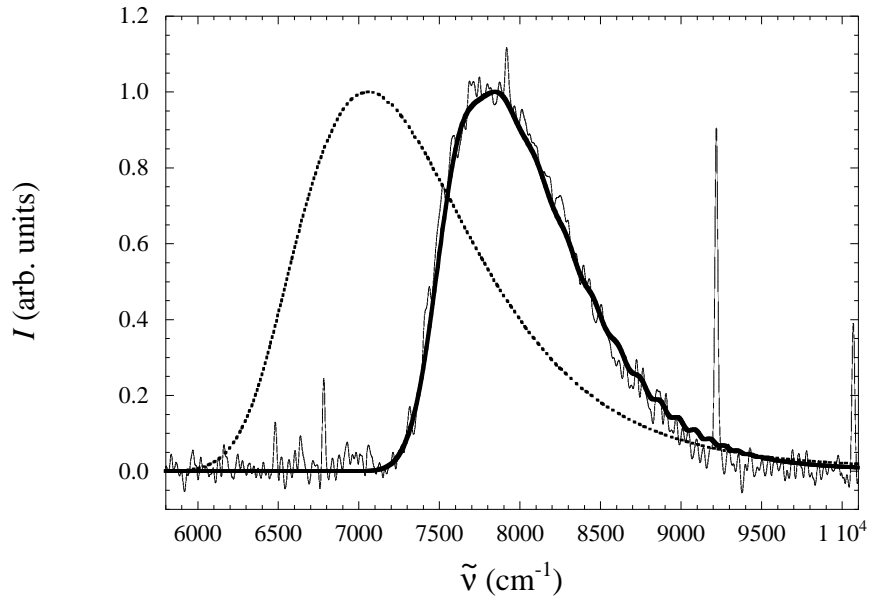}
\caption{ Comparison of the experimental and calculated spectra. Dashed curve: spectrum computed by using literature potentials. Thick solid line: spectrum computed by using a $R^{-12}$ repulsive potential with adjustable parameters. \label{fig:simspc}}\end{center}\end{figure}

The overlap integrals in Eq. \ref{eq:FCspc} are evaluated by spline interpolation and quadrature \cite{champagne}. Moreover, the calculated spectrum has been convoluted with the instrumental function to account for the finite resolution of 16 cm$^{-1}.$
In Fig. \ref{fig:simspc} we compare the experimental and calculated spectra.

The dashed line is the spectrum calculated by using the theoretical potential energy curves found in literature. The agreement is only qualitative. The synthetic spectrum is centered at much smaller wavenumbers than the experimental one and also its width is much larger than observed. These features are the consequence of the fact that the theoretical potential energy curves are too close to each other and that the dissociative potential is probably too steep in the region of the minimum of the bound state.
The only correct feature shown by the computed spectrum is its asymmetric shape with a longer tail towards higher energies. This feature is due to the non negligible contribution of vibrational states with $v^{\prime}>0.$

It is, however, possible, by inverting the line shape, to reconstruct the repulsive part of the dissociative potential in a coordinate range, over which the vibrational wavefunctions of the bound state are non negligible \cite{telli}. 

As only few vibrational levels of the bound state are thermally excited, the choice of the analytical form of the bound state potential is not too critical and we assume that the bound state is correctly described by the theoretical literature potential $(3)0_{u}^{+}$~\cite{jonin2002_II}. We also assume that the repulsive part of the potential of the dissociative $(1)0_{g}^{+}$ state is described by an inverse power law potential
of the type
\begin{equation}
V_{r} = A +\frac{V_{0}}{x^{12}}\qquad\left( x={R}/{R_{e^{\prime}}}\right)
\label{eq:vrep}\end{equation}
 in which $A$ and $V_{0}$ are adjustable parameters. For each pair $(A,\, V_{0})$ the scattering eigenfunctions are to be computed anew in order to calculate the Franck--Condon factors. The best agreement with the observed spectrum is obtained by setting $A=(5315\pm 32)$ cm$^{-1}$ and $V_{0} = (760\pm 16)$ cm$^{-1}.$ The spectrum computed by using these values of the parameters is shown as a thick solid line in Fig. \ref{fig:simspc} and is in nearly perfect agreement with the experimental one. The blue asymmetry of the experimental spectrum is also well reproduced and is due to the contributions of the vibrational states with $v^{\prime}>0.$ 

\begin{figure}[htbp]
 \begin{center}
 \includegraphics[scale=0.5]{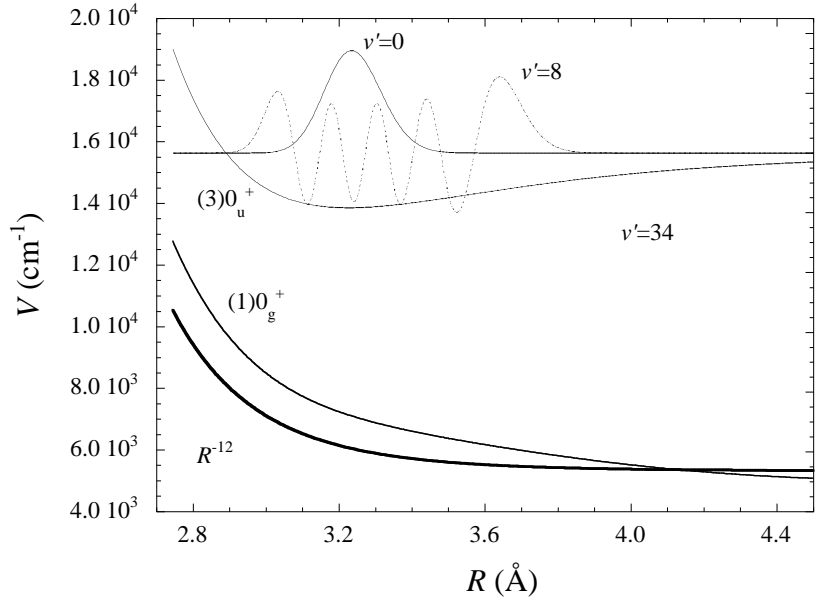}
\caption{Comparison of the potential energy curves for the dissociative state $0_{g}^{+}$. Thin solid line: literature potential \cite{jonin2002_II}. Thick solid line: best fit potential. The upper bound state potential $(3)0_{u}^{+}$ is reported along with two vibrational eigenfunctions for $v^{\prime} =0$ and 8. \label{fig:potvibf}}\end{center} 
 \end{figure}
\noindent In Fig. \ref{fig:potvibf} we compare the theoretical and best fit potentials. We also plot the potential energy curve of the upper bound state along with the vibrational eigenfunctions for $v^{\prime} =0 $ and 8, respectively, in order to show over what coordinate range the comparison between the two determinations of the dissociative potential is meaningful. 

The difference between the theoretical and best fit potentials (approximately 880 cm$^{-1}$ for $R= R_{e^{\prime}}$) is fairly large. Probably, a theoretical determination of the electronic transition moments for these states or an improvement of the theoretical calculations of the potential energy curves might reduce the disagreement between the experimental and theoretical determination of the potential energy curves.

In any case, it has been demonstrated that the IR emission spectrum is due to a transition between two molecular states correlated with atomic limits in the $6p$ and $6s$ manifold, respectively.

\section{High--density results}
When the density $N$ of the gas is increased, the increasingly stronger interaction of the Xe$_{2}$ excimers with the environment affects the features of the emission spectrum. On one hand, as expected, the collision frequency increase reduces the lifetime of this species. Thus, the spectrum width broadens linearly with $N.$

 \begin{figure}[htbp]
 \begin{center}
 \includegraphics[scale=0.5]{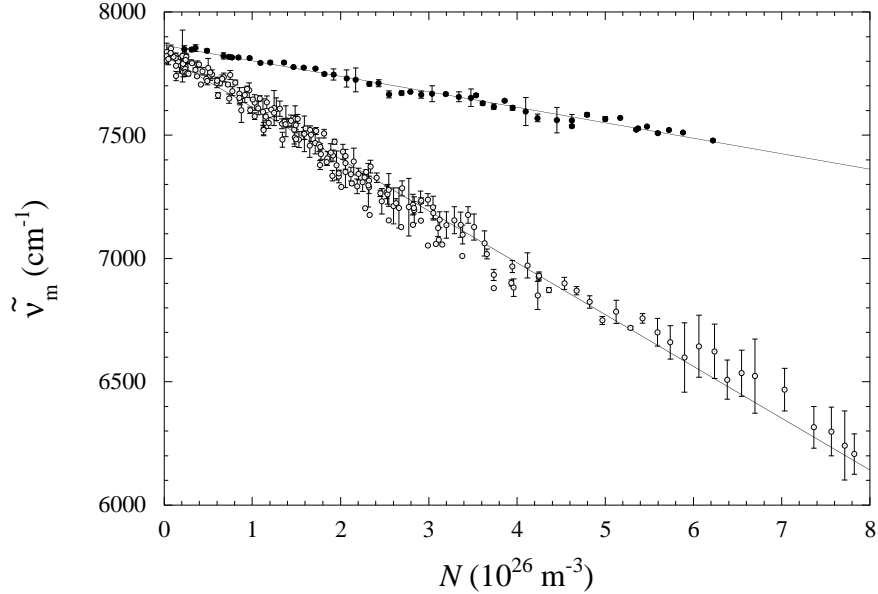}
\caption{Density--dependent shift of the center of the excimer emission spectrum. Open dots: pure Xe. Closed dots: Ar(90\%)--Xe(10\%) mixture. Lines: prediction of the theoretical model. \label{fig:shift}}\end{center} 
 \end{figure}
 \noindent On the other hand, the center of the emission band shifts unexpectedly to lower wave numbers as $N$ increases. A linear relationship of the type \begin{equation}
\tilde\nu_{m} = \tilde \nu_{m,0}-AN
\label{eq:linshift}\end{equation}
describes accurately the experimental data. 

This linear trend is observed both in pure Xe gas and in the mixture, although the slope has quite different values in the two different gases, and is obeyed for densities as high as $N\approx 8\cdot 10^{26}$ m$^{-3},$ i. e., for densities nearly 40 times larger than the density of the ideal gas at $T=273$ K and $P=0.1$ MPa.

In Fig. \ref{fig:shift} the central wave number of the spectrum, $\tilde\nu_{m},$ is shown as a function of $N.$ In the limit $N\rightarrow 0,$ $\tilde \nu_{m}$ has the same value, the Xe$_{2}$ excimer being produced either in the pure gas or in the mixture. 

By contrast, when the Xe$_{2}$ excimer is produced in the pure gas,  the density dependence of $\tilde \nu_{m}$ is much stronger than in the case in which Xe$_{2}$ is produced in the mixture.  If the data are fit to straight lines, the following values for the slope are obtained: $A = (2.08 \pm 0.04)\cdot 10^{-22}$ $m^{2}$ in pure Xe and $A=(0.63\pm0.015)\cdot 10^{-22}$ m$^{-2}$ in the mixture. We have to stress the fact, however, that the lines in Fig. \ref{fig:shift} are not such linear fits but represent the predictions of a simple theoretical model that we have developed in order to understand the physics of the phenomenon we are investigating \cite{borghesani2001}.

According to the suggestions of Mulliken \cite{mulliken1970}, an excited state of a diatomic rare gas molecule can be envisaged as an ionic core, e. g., Xe$_{2}^{+},$ plus a Rydberg electron in an extended orbital. We assume that the electron is so largely delocalized that many atoms of the host gas are encompassed within the large electron orbit. These atoms are polarized by the strong field of the ionic core. If the density is large enough, we can assume that the dielectric screening exerted by the polarized atoms can be treated by introducing the density--dependent dielectric constant of the gas, $K(N).$ 

The interaction potentials are of electrostatic nature and are therefore reduced by a factor $1/K^{2}$ with respect to the dilute gas condition. At the densities of the experiment, the dielectric constant can be Taylor--expanded to first order as a function of $N:$ $K\approx 1+ N\alpha/\epsilon_{0},$ where $\alpha $ is the atomic polarizability of the atoms of the host gas and $\epsilon_{0}$ it the vacuum permittivity.

The energy difference between the upper bound potential $V_{0_{u}^{+}}$ and the lower dissociative potential $V_{0_{g}^{+}}$ is reduced by the same factor. 
The average energy difference that approximately corresponds to the central wave number of the observed emission spectrum is thus given by
\begin{equation}
\tilde\nu_{m} = \frac{\langle V_{0_{u}^{+}} - V_{0_{g}^{+}}\rangle}{hcK^{2}\left(N\right)}\approx \tilde\nu_{m,0} \left(1 - \frac{2\alpha}{\epsilon_{0}}N\right)
\label{eq:d1}\end{equation}
in which $\tilde\nu_{m,0}=\langle  V_{0_{u}^{+}} - V_{0_{g}^{+}} \rangle/hc$ and $\langle \ldots\rangle$ is a suitable weighted average of the potentials over the coordinates of the vibrational states of the bound potentials. $\tilde\nu_{m,0}$ is the zero--density limit of the central wave number of the emission spectrum. 
A linear dependence of $\tilde \nu_{m}$ is thus obtained.

Unfortunately, if only this solvation effect were accounted for, the slope $\tilde\nu_{m,0}\alpha/\epsilon_{0}$ would amount to only half of the observed value. The missing contribution arises from quantum many--body effects in the interaction of the delocalized electron with the atoms of the surrounding gas. In fact, the wavelength of the loosely bound Rydberg electron in the excited dimer is very large and spans over many surroundings atoms. Such a state can exist provided that the electron is weakly scattered off the gas atoms, i. e., if the electron mean free path is much larger than the radius of its orbit, as it occurs in the present case \cite{borghesani2001}. 

In this situation, the Rydberg electron can be considered as nearly free and its simultaneous interaction with many atoms of the host gas leads to a density--dependent shift of its energy given by \cite{fermi}
\begin{equation}
\Delta E (N) =  \frac{2\pi \hbar^{2}}{m} Na
\label{eq:fermi}\end{equation}
where $a$ is the electron--atom scattering length and $m$ is the electron mass.

This contribution has to be added to the energy difference between the initial and final state of the molecule, eventually leading to the following expression for the energy of the emitted IR photon in the center of the spectrum
\begin{equation}
\tilde\nu_{m} =  \tilde\nu_{m,0} \left(1 - \frac{2\alpha}{\epsilon_{0}}N\right) + \frac{\hbar}{mc}Na =  \tilde\nu_{m,0} -\left( \tilde\nu_{m,0}\frac{2\alpha}{\epsilon_{0}} -\frac{\hbar a}{mc}\right)N
\label{eq:wavenumb}\end{equation}
The present model predicts a linear density--dependence of the wave number corresponding to the maximum of the spectrum, as actually observed. Moreover, the slope can be calculated if the gas parameters $\alpha$ and $a$ are known. 
In the case of the heavier rare gases, the scattering length is negative and its contribution effectively adds to the solvation contribution.

In the pure gas, the choice of the gas parameters is obvious. In the case of the mixture, the low concentration of Xe leads to the assumption that the excimer is actually surrounded by Ar atoms so that the parameters of Ar have to be chosen.

In the case of the pure Xe gas, $\alpha = 4.45\cdot 10^{-45}$ F$\cdot $m \cite{mait} and $a=-3.09$ \AA\ \cite{zecca}, yielding $\left({2\tilde\nu_{m,0}\alpha}/{\epsilon_{0}} -{\hbar a}/{mc}\right) = 1.98 \cdot10^{-22}$ m$^{2},$ to be compared with the experimental value $(2.08 \pm 0.04)\cdot 10^{-22}$ $m^{2}.$ In the cas of the mixture, the Ar parameters are $\alpha = 1.827\cdot 10^{-40}$ F$\cdot$m \cite{mait} and $a =-0.86 $ \AA\ \cite{zecca}, yielding a slope value $0.66 \cdot 10^{-22}$ m$^{2},$ to be compared with the experimental value $(0.63\pm0.015)\cdot 10^{-22}$ m$^{-2}.$ 

The theoretical prediction agrees very well with the experiment. In Fig \ref{fig:shift} the straight lines actually represent the theoretical prediction and are not a mere linear fit to the data. 

It is interesting to note that the dielectric screening effect always produces a red--shift of the excimer emission spectrum because it reduces the strength of the Coulomb interaction between the Rydberg electron and the ion core. On the contrary, 
the multiple scattering effect gives a contribution to the energy of the emitted photon whose sign depends on the nature of the (nearly free) electron--atom interaction. If this is attractive, as in the case of the heavier rare gases, the electron--atom scattering length is negative, whereas it is positive in the case of Ne and He. Thus, roughly speaking, the density--dependent shift of the emission spectrum might be ``fine tuned'' from red-- to (even) blue--shift by suitably choosing the buffer gas in which Xe excimers are produced. Further investigations in different mixtures should help confirming this model.

\section{Conclusions}
The study of the infrared emission spectrum of Xe excimers in pure Xe gas or in an Ar--Xe mixture reveals interesting features in both the low-- and high--density limits. 
The analysis of the spectrum at low density has allowed the assignment of the molecular states involved in the transition. This was never done before because no IR molecular spectra of Xe in the present range were observed before. Moreover, it has been possible to give an experimental determination of the potential energy curve of a gerade molecular state related to the $6s$ manifold. 

On the other hand, it has been possible to see how cooperative-- and many--body effects continuously develop by increasing the density up to quite high values. The change of the excimer emission spectrum with the gas density reveals that (at least) two kinds of mechanisms, one classical and one quantum, affect the interaction of the excimer with its environment. In our opinion, it is worth noting that the quantum contribution to the self--energy of the nearly free Rydberg electron links a typical issue of molecular physics such as the emission spectrum of a decaying diatomic molecule to the (apparently) uncorrelated realm of electron transport in dense gases. In fact, also in that case the density dependent shift of the electron energy due to multiple scattering deeply affects electron transport \cite{borg1988,borg1992,borg2002}.

\bibliography{MOLECXVI}
\end{document}